\documentclass{aastex631} 
\usepackage{amsmath}
\newcommand{\R}{\mathbb{R}}
\newcommand{\mitkavli}{MIT Kavli Institute for Astrophysics and Space Research, Massachusetts Institute of Technology, 77 Massachusetts Ave, Cambridge, MA 02139, USA}
\newcommand{\mitphysics}{Department of Physics, Massachusetts Institute of Technology, 77 Massachusetts Ave, Cambridge, MA 02139, USA}

\begin{document}

\title{Constraining the Faint-End Slope of the FRB Energy Function Using CHIME/FRB Catalog-1 and Local Volume Galaxies} 
\shorttitle{}
\shortauthors{}

\author[0000-0002-3615-3514]{Mohit~Bhardwaj}
  \affiliation{McWilliams Center for Cosmology, Department of Physics, Carnegie Mellon University, Pittsburgh, PA 15213, USA}

  \author[0000-0001-9345-0307]{Victoria M. Kaspi}
\affiliation{Department of Physics, McGill University, 3600 rue University, Montr\'eal, QC H3A 2T8, Canada}
\affiliation{Trottier Space Institute, McGill University, 3550 rue University, Montr\'eal, QC H3A 2A7, Canada}

\author[0000-0002-4279-6946]{K.~W.~Masui}
\affiliation{\mitkavli}
\affiliation{\mitphysics}

\author[0000-0002-3382-9558]{B.~M.~Gaensler}\affiliation{Dunlap Institute for Astronomy \& Astrophysics, University of Toronto, 50 St.~George Street, Toronto, ON M5S 3H4, Canada}
\affiliation{David A.~Dunlap Department of Astronomy \& Astrophysics, University of Toronto, 50 St.~George Street, Toronto, ON M5S 3H4, Canada}
\affiliation{Department of Astronomy and Astrophysics, University of California Santa Cruz, 1156 High Street, Santa Cruz, CA 95064, USA}

\author[0000-0003-2405-2967]{Adaeze L.~Ibik}
  \affiliation{Dunlap Institute for Astronomy \& Astrophysics, University of Toronto, 50 St.~George Street, Toronto, ON M5S 3H4, Canada}
  \affiliation{David A.~Dunlap Department of Astronomy \& Astrophysics, University of Toronto, 50 St.~George Street, Toronto, ON M5S 3H4, Canada}

  \author[0000-0002-4623-5329]{Mawson W. Sammons}
\affiliation{Department of Physics, McGill University, 3600 rue University, Montr\'eal, QC H3A 2T8, Canada}
\affiliation{Trottier Space Institute, McGill University, 3550 rue University, Montr\'eal, QC H3A 2A7, Canada}

\correspondingauthor{Mohit Bhardwaj}
\email{mohitb@andrew.cmu.edu}

\begin{abstract}
 Despite hundreds of detected fast radio bursts (FRBs), the faint-end slope ($\gamma$) of their energy distribution remains poorly constrained, hindering understanding of whether bright, cosmological FRBs and faint, Galactic magnetar SGR~1935+2154-like bursts share a common origin. In this study, we constrain this faint-end slope, modeled with a Schechter-like distribution, by searching for potential associations between bursts from the CHIME/FRB Catalog-1 and galaxies in the local volume. We cross-matched Catalog-1 FRBs with 495 local volume galaxies within 21 Mpc, identified from the HECATE catalog, and found no associations. Assuming the FRB energy function extends to $\sim 3 \times 10^{34}$ erg—the energy of the Galactic magnetar burst from SGR~1935+2154—this null result constrains \(\gamma\) to be $<$ 2.3 (95\% confidence upper limit), representing the first empirical estimate for extragalactic FRBs at such low energies. This finding supports the hypothesis that the FRB population is dominated by bright, likely cosmological bursts with a relatively flat energy distribution ($\gamma < 2.5$). However, the constraint weakens if higher energy thresholds are assumed. A flatter energy function is consistent with the observed anti-correlation between FRB dispersion measure and fluence, as seen across various observational bands. While the contribution of low-energy bursts, such as those from the Galactic magnetar SGR 1935+2154, appears minimal, our results suggest that normal magnetars like SGR 1935+2154 could dominate the FRB population if their burst rates and energies scale with age and magnetic field. Upcoming CHIME/FRB Catalog-2 data and targeted nearby galaxy surveys will further refine these constraints, offering critical insight into whether FRBs arise from a single population or diverse origins.
\end{abstract}


\section{Introduction}
\label{sec:intro}

Fast Radio Bursts (FRBs) are a class of highly energetic astrophysical transients characterized by brief pulses of radio emission lasting only $\sim$milliseconds \citep{lbm+07, tsb+13}. Since their discovery, FRBs have been detected across cosmological distances \citep{2023Sci...382..294R}, making them a powerful tool for probing the distribution of ionized baryonic matter throughout the Universe \citep{Macquart2020Nature}. Despite the detection of nearly 1,000 FRBs to date\footnote{For a complete list of known FRBs, see the TNS: \url{https://www.wis-tns.org/} \citep{2020TNSAN..70....1Y}.}, their origins remain largely unknown. Furthermore, the wide range of burst properties—encompassing a broad range of spectro-temporal and polarimetric characteristics, along with the diversity of host environments—suggests a complex and heterogeneous set of progenitors and/or emission mechanisms \citep{2022A&ARv..30....2P,2023RvMP...95c5005Z}. 

Current observations indicate that a substantial fraction of FRBs repeat, implying that their progenitors are capable of producing multiple bursts over time and thus have a non-cataclysmic origin \citep{2019NatAs...3..928R, Bhardwaj2021B,2021MNRAS.500.3275C,2023PASA...40...57J}. Magnetars, neutron stars with exceptionally strong magnetic fields \citep[$\gtrsim 10^{14}$ G;][]{1992ApJ...392L...9D,2017ARA&A..55..261K}, are considered strong candidates for these repeating FRBs. This view is supported by the detection of FRB-like bursts \citep{CHIME2020Nature,Bochenek2020Nature} from the young Galactic magnetar SGR 1935+2154 \citep[spin-down age $\sim$ 3.6 kyr;][]{2016MNRAS.457.3448I}. However, the discovery of repeating FRBs 20200120E and FRB 20240209A in a globular cluster of the nearby spiral galaxy M81 \citep{Bhardwaj2021A,Kristen2022Nature} and in the outskirts of a quiescent elliptical galaxy \citep{2025ApJ...979L..21S,2025ApJ...979L..22E}, respectively, challenges the young progenitor hypothesis, suggesting that some FRBs may originate from older stellar populations.
Despite these challenges, more recent studies propose that the majority of FRB sources likely originate in galactic disks \citep{2024arXiv240801876B}. Additionally, the prevalence of FRBs in late-type galaxies in the local Universe \citep{2023ApJ...954...80G,2024ApJ...971L..51B} supports young progenitor models, such as core-collapse supernovae (CCSNe), as a primary channel for FRB production. However, other progenitor channels may contribute, and the specific subtypes of CCSNe most likely responsible for FRBs remain uncertain \citep{2020ApJ...905L..30S,2020MNRAS.498.3927H,2020ApJ...899L...6L,2022ApJ...924L..14Z,2024ApJ...969..123L,2024arXiv240603672C,2024arXiv240600476Z,2024arXiv240506281W,2024arXiv240916964S}.

Population studies of FRBs detected by wide-field surveys have been crucial to advance our understanding of their origins. These studies enable statistical analysis of event rates and energy distributions, providing important constraints on the evolutionary history of FRB progenitors \citep{2018MNRAS.474.1900M, Luo2018MNRAS, 2019ApJ...883...40L,Luo2020MNRAS, Hashimoto2020MNRASa, 2020MNRAS.498.3927H, Zhang2021MNRAS, 2022MNRAS.509.4775J, 2023ApJ...944..105S, 2024Univ...10..207Z,2025arXiv250109810G}. Targeted studies, on the other hand, by focusing on specific galaxy populations, such as starburst galaxies, nearby galaxy clusters, or sites of various transient events, offer novel ways to refine and test FRB progenitor models \citep{fialkov2018ApJ, agrawal2019MNRAS,2020MNRAS.493.5170H,2021MNRAS.501..541P,2023A&A...674A.223P,2024MNRAS.528.6340P,2024MNRAS.tmp.1841H,2025arXiv250601238L}. In this context, local volume galaxies present a valuable opportunity to identify FRBs that may be suitable candidates for multi-wavelength counterpart studies, particularly given the sensitivity limitations of high-energy telescopes \citep{2023arXiv230810930P,2024arXiv240811895C}. Targeted observations of FRBs from these nearby galaxies can also help constrain the faint-end slope ($\gamma$) of the FRB energy function at low energies (\(\lesssim 10^{39}\) erg), a challenging task to achieve with wide-field untargeted studies published to date \citep[for instance,][]{2023ApJ...944..105S,2024arXiv240809351A}. This approach can also enable the assessment of whether an intrinsic lower energy cutoff exists \citep{2016MNRAS.461L.122L,2020ApJ...891...82W}, distinguishing FRBs from the Galactic neutron star population, which produces bursts with some similar phenomenological characteristics but at much lower energies.

In this study, we constrain $\gamma$ by exploring potential associations between fast radio bursts (FRBs) detected in the first catalog released by the CHIME/FRB collaboration and galaxies within the local volume. We define the ``local volume" as the region containing galaxies within a luminosity distance of $\lesssim$ 21 Mpc, equivalent to the distance of the second closest extragalactic FRB discovered. The CHIME/FRB Catalog-1, which spans the period from 2018 July 25 to 2019 July 2, includes 536 FRBs: 62 bursts from 18 repeating sources and 474 one-off events \citep{CHIMEFRBcatalog}. As the largest sample of FRBs from a single wide-survey instrument \citep[sensitive to $\sim$ 2/3 of the entire sky;][]{2018ApJ...863...48C} published to date, this catalog provides a unique opportunity to use possible connections between FRBs and nearby galaxies to better understand the FRB energy distribution.

The structure of this paper is as follows: In \S\ref{sec:formalism}, we outline the model and formalism used to constrain the FRB population. In \S\ref{sec:LVsample}, we describe the local volume galaxy sample selected for this study. In \S\ref{sec:results}, we detail the methodology used to search for associations between FRBs and selected galaxies and present the results of our analysis. Finally, the broader implications of our findings are discussed in \S\ref{sec:discussion}, with conclusions presented in \S\ref{sec:conclude}. Throughout this study, we adopt Planck cosmological parameters \citep{2020A&A...641A...6P}.

\section{FRB Population Modeling: Formalism}
\label{sec:formalism}

In this section, we outline the formalism used to model the observed FRB population. We investigate two primary scenarios: one in which the FRB rate is proportional to the star formation rate (SFR) of the host galaxy, and another where it is proportional to the stellar mass (SM). These two models offer distinct hypotheses regarding the origin of FRB sources.

The \textbf{SFR model} posits that FRBs are linked to young stellar populations, such as massive stars or high-mass X-ray binaries, commonly associated with star-forming regions \citep{Bochenek2021ApJ, James2022MNRASsfr}. This model is particularly relevant if most FRBs originate from young magnetars, such as Galactic magnetar SGR 1935+2154 \citep{2016MNRAS.457.3448I}, formed through CCSNe.

In contrast, the \textbf{SM model} suggests that FRBs are predominantly associated with older stellar populations, such as those found in globular clusters \citep{2013ApJ...772...82H, 2013MNRAS.429.2881C} or low-mass X-ray binaries \citep{2004MNRAS.349..146G}, where the FRB rate is proportional to the stellar mass of the host galaxy \citep{2022ApJ...924L..14Z,2023ApJ...944....6K}. This model is more suitable to explain the source of FRB 20200120E which has been linked to a globular cluster in M81.

To model the FRB rate above a certain energy threshold for a galaxy, we assume the following relation:

\begin{equation}
R_{\text{gal}}(E \geq E_{\text{th}}) = K_{0} \times \left( \frac{\text{SFR or SM}}{\text{SFR}_{\text{MW}}~\text{or}~\text{SM}_{\text{MW}}} \right) \int_{E_{\text{th}}}^{\infty} \frac{1}{E_*} \left( \frac{E}{E_*} \right)^{-\gamma} \exp\left( -\frac{E}{E_*} \right) dE,
\label{Eq1:gal_rate}
\end{equation}

\noindent where \( K_{0} \) is the proportionality constant for each model with units consistent with $R_{\text{gal}}$($E \geq E_{\text{th}}$) which is day$^{-1}$ in our case, \( E_{\text{th}} \) is the minimum intrinsic energy threshold for FRBs, \( \gamma \) represents the faint-end slope of the differential FRB energy function and $E_*$ denotes the maximum cutoff energy, beyond which the energy function exhibits an exponential decline. The rates are normalized relative to the Milky Way (MW), using its SFR = 1.65 M$_{\odot}$yr$^{-1}$ and SM = $6.08 \times 10^{10}$ M$_{\odot}$ as estimated by \cite{2015ApJ...806...96L}. In this analysis, we assume that the FRB population follows a Schechter-like energy function, which is widely used to model the luminosity and energy distributions of extragalactic sources \citep{1976ApJ...203..297S}, and adopted in most FRB population studies \citep[e.g.,][]{Chawla2021,2022MNRAS.509.4775J,2023ApJ...944..105S}. This choice ensures consistency with the published constraints shown in \S\ref{sec:discussion}, which all employ the same form. Although alternatives such as broken power laws or log-normal distributions have been proposed \citep[e.g.,][]{2022Ap&SS.367...66C,2024ApJ...973L..54C}, they are not yet observationally well constrained and would hinder direct comparison with existing results.

Next, we derive $K_{0}$ by fitting the observed all-sky FRB rate as estimated by \cite{CHIMEFRBcatalog} at 600 MHz. To accomplish this, we integrate the FRB rate per galaxy, given by Equation \ref{Eq1:gal_rate}, over the cosmic star formation history or cosmic stellar mass density, depending on the model. The total all-sky FRB rate, $\R_{\text{all-sky}}$, is given by:

\textbf{SFR Model:}
\begin{equation}
\R_{\text{all-sky}} = \int_{0}^{z_{\text{max}}} dz \, \frac{dV}{dz} \, \frac{K_{0} \times \Phi(z)}{(1+z) \times \text{SFR}_{\text{MW}}} \int_{E_{\text{lim}}}^{\infty} \frac{1}{E_*} \left( \frac{E}{E_*} \right)^{-\gamma} \exp\left( -\frac{E}{E_*} \right) dE,
\label{Eq2:rate_sfr}
\end{equation}

\textbf{SM Model:}
\begin{equation}
\R_{\text{all-sky}} = \int_{0}^{z_{\text{max}}} dz \, \frac{dV}{dz} \, \frac{K_{0} \times \rho(z)}{(1+z) \times \text{SM}_{\text{MW}}} \int_{E_{\text{lim}}}^{\infty} \frac{1}{E_*} \left( \frac{E}{E_*} \right)^{-\gamma} \exp\left( -\frac{E}{E_*} \right) dE,
\label{Eq3:rate_sm}
\end{equation}

where:
\begin{itemize}
    \item \( \Phi(z) \) represents the cosmic star formation rate density in the SFR model \citep{2014ARA&A..52..415M}; see Equation \ref{appendix:EqA2_SFRD}.
    \item \( \rho(z) \) denotes the cosmic stellar mass density in the SM model, as detailed in Appendix \ref{appendix:sm}.
    \item \( \frac{dV}{dz} \) is the differential comoving volume element.
    \item The maximum redshift limit, $z_{\rm max}$, represents the redshift range over which the adopted all-sky FRB rate is sensitive. We adopt $z_{\rm max} = 2$, consistent with the 95\% credible upper limit of the CHIME/FRB redshift distribution derived by \cite{2023ApJ...944..105S}. This choice accounts for potential incompleteness in the FRB distribution at higher redshifts due to propagation effects \citep{2022ApJ...934...71O, 2024Natur.634.1065B}, which are not considered in Equations \ref{Eq2:rate_sfr} and \ref{Eq3:rate_sm}. However, as demonstrated in Appendix \ref{appendix:rate_zmax}, increasing $z_{\rm max}$ does not have any meaningful impact on  our resulting $\gamma$ constraint. 
    
\end{itemize}

\begin{figure}[h]%
    \centering
    \includegraphics[origin=top, width=.95\linewidth]{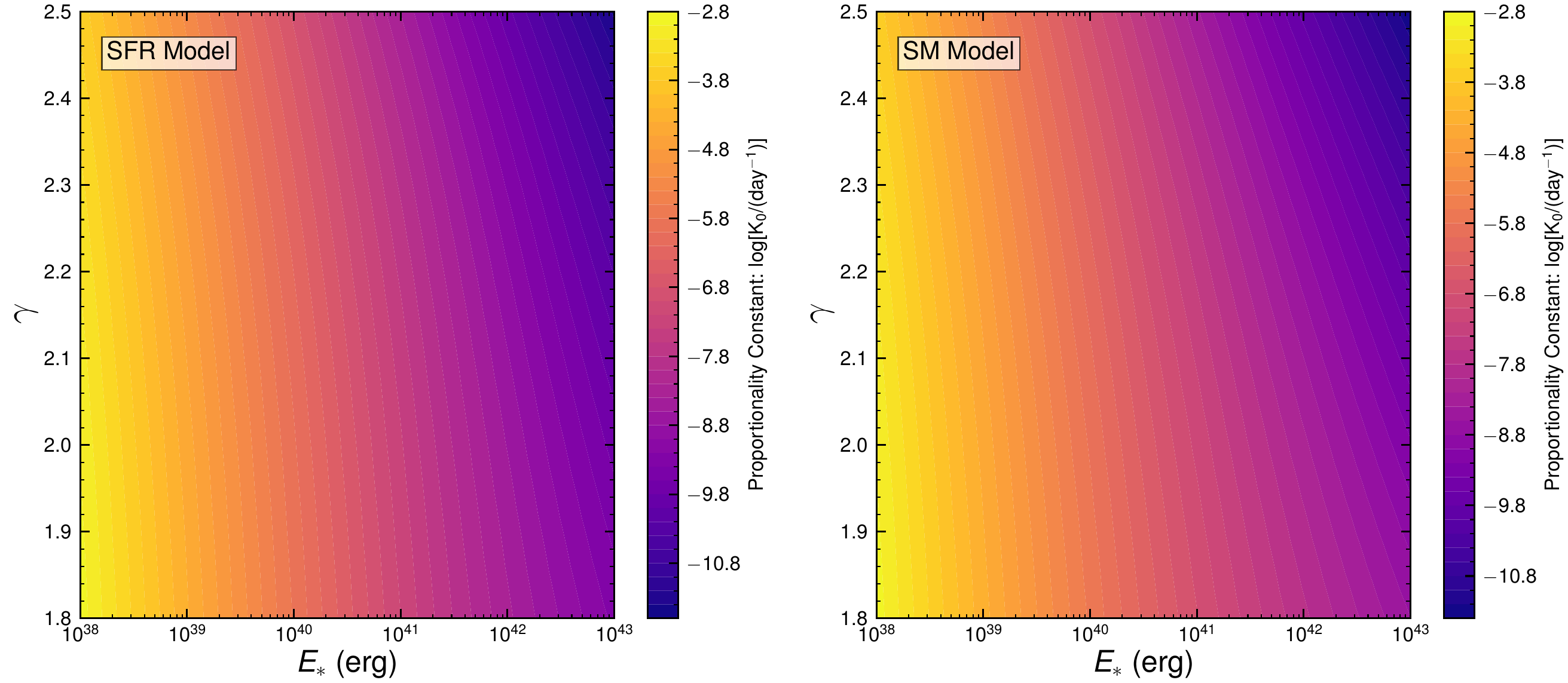}%
    \caption{The estimated proportionality constant $K_{0}$ as a function of $\gamma$ and $E_{*}$ for the SFR (left) and SM (right) models. The all-sky rate $\R_{\text{all}-\text{sky}} = 525$ day$^{-1}$ was used in Equations \ref{Eq2:rate_sfr} and \ref{Eq3:rate_sm}, assuming $z_{\rm max} = 2$. Note that $K_{0}$ is scaled using the Milky Way SFR and stellar mass. For more details, refer to \S\ref{sec:formalism}.}
    \label{fig:Ko}%
\end{figure}

The term \( (1+z)^{-1} \) accounts for cosmological time dilation. The limits of the inner integral are set by the energy threshold \( E_{\text{lim}} \), computed as:

\begin{equation}
E_{\text{lim}} = \max\left(E_{\text{th}}, \frac{4\pi D_L(z)^2 F_{\nu} \delta\nu}{(1+z)^{2+\alpha}} \right),
\label{Eq:Elim}
\end{equation}

\noindent where \( F_{\nu} \) is the fluence threshold of the radio telescope, \( \delta\nu \) is the bandwidth, and \( D_L(z) \) is the luminosity distance at redshift \( z \). In this study, we adopt \( F_{\nu} = 5 \, \text{Jy ms} \), which is the pivot fluence above which the CHIME/FRB all-sky rate is best constrained \citep{CHIMEFRBcatalog}, and \( \Delta\nu = 0.4 \, \text{GHz} \), the bandwidth of the CHIME telescope. When converting fluence to energy, we assume the spectral index $\alpha = -1.39$ from \cite{2023ApJ...944..105S}. We also varied the parameter $\alpha$ over the range $[0, -2]$ and found no significant impact on our results, as demonstrated in Appendix \ref{appendix:alpha}. For this analysis, we set \( E_{\text{th}} = 3 \times 10^{34} \) erg, based on the energy of the FRB-like burst from SGR 1935+2154 detected by \cite{CHIME2020Nature}. In Appendix \ref{appendix:Eth}, we discuss the effect of varying $E_{\text{th}}$ on the derived $\gamma$ constraint.
Note that in scenarios where the formation of FRB sources is delayed relative to the starburst epoch in galaxies \citep{2024ApJ...969..123L}, the FRB rate is likely influenced by both the host galaxy's SFR and SM, as observed in other astrophysical transients that exhibit time delays with respect to the cosmic star formation history, such as short gamma-ray bursts \citep{2022ApJ...940L..18Z} and Type Ia supernovae \citep{2014ARA&A..52..107M}. In these cases, we expect the \( K_{0} \) value to lie between the limits derived from the SFR and SM models.

Using the CHIME/FRB all-sky rate of 525 bursts per day for fluences above 5 Jy ms at 600 MHz, with scattering times under 10 ms and dispersion measures (DM) above 100 pc cm$^{-3}$ \citep{CHIMEFRBcatalog}, we estimate \( K_{0} \) as a function of \( \gamma \) and \( E_{*} \) for both models, as shown in Figure \ref{fig:Ko}. These estimates of \( K_{0}(\gamma, E_{*}) \), combined with Equation \ref{Eq1:gal_rate}, are then used to constrain \( \gamma \) using the local volume galaxy sample, as discussed in subsequent sections.

\section{Local Volume Galaxy Selection}
\label{sec:LVsample}
\begin{figure}[ht]
\begin{center}
\includegraphics[width=1\linewidth]{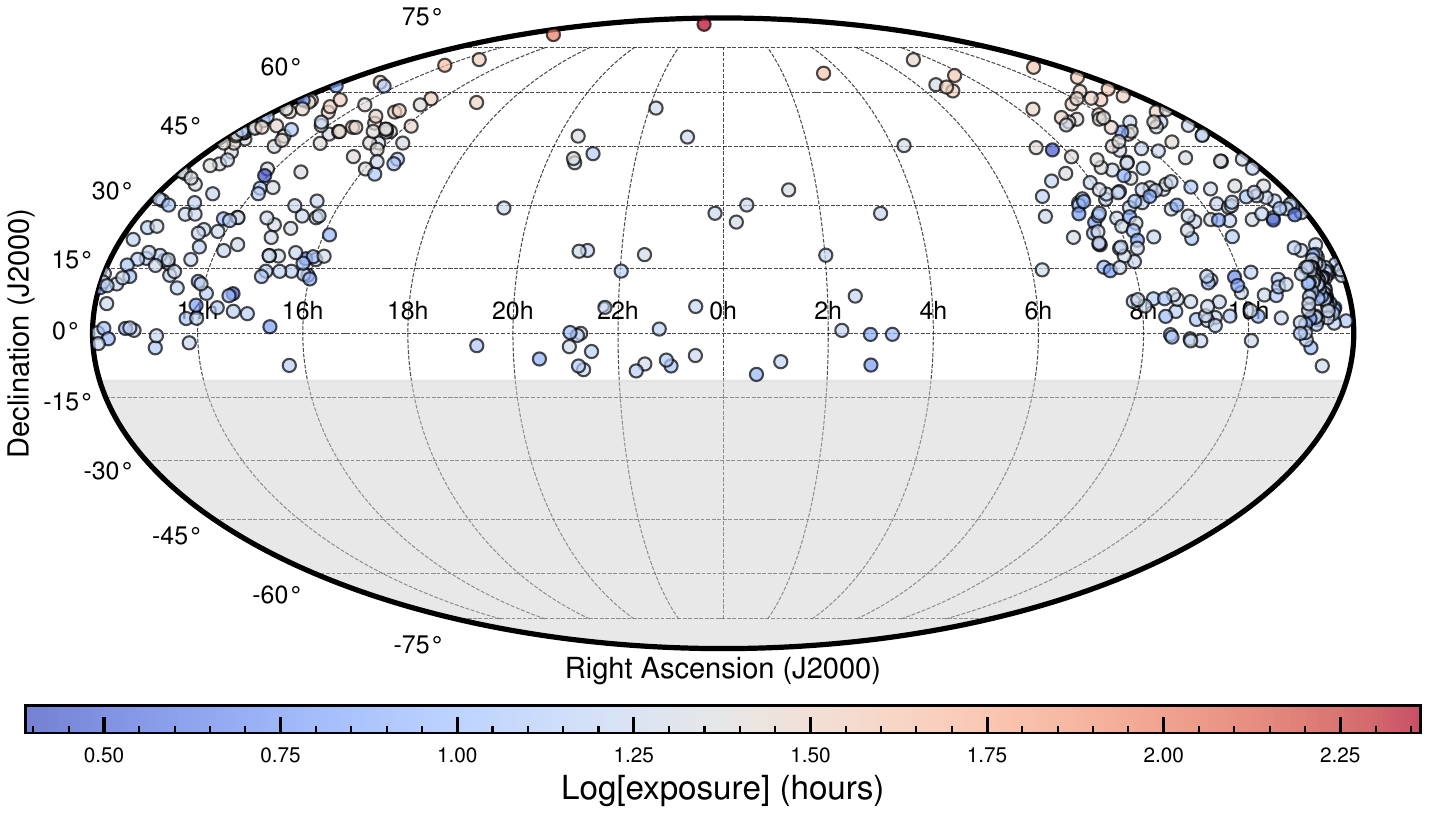}
\end{center}
\caption{Sky distribution (Mollweide projection) of local Universe galaxies selected from the HECATE catalog \citep{2021MNRAS.506.1896K}. The galaxies are represented by circles, color-coded based on their mean exposure in CHIME/FRB Catalog 1. The shaded light gray region indicates the area below CHIME/FRB’s field of view \citep[$\delta < -11^{\circ}$;][]{2018ApJ...863...48C}.}
\label{fig:frbfov}
\end{figure}

In order to constrain the FRB energy function, we search for potential associations between Catalog 1 FRBs and local volume galaxies, identified using the HECATE catalog \citep{2021MNRAS.506.1896K}. The HECATE catalog is a comprehensive resource, containing detailed physical properties, such as inclination angle, size, distance, SFR, and SM, for 204,733 galaxies in the local Universe (luminosity distance $<$ 200 Mpc). In our work, we focus on local volume galaxies within a luminosity distance of 21 Mpc, based on several considerations. First, this distance encompasses the host of the second-closest extragalactic FRB 20181030A, NGC 3252 \citep{Bhardwaj2021B}, which according to the HECATE catalog is located at 20.8 Mpc. This distance limit enables us to place an informative constraint on $\gamma$ at $E_{\rm th} \approx 10^{36}$ erg, corresponding to a fluence limit of 5 Jy ms at 21 Mpc (see Section~\ref{subsec:gamma-r4}). We note that the a posteriori selection of this distance does not introduce bias into our analysis, as demonstrated via simulation in Appendix~\ref{appendix:sim}.
Additionally, the relatively small co-moving volume within 21 Mpc and the low galaxy number density can permit the identification of potential associations with a low chance probability ($<$ 10\%). This estimate does not account for the look-elsewhere effect—the increased likelihood of obtaining a seemingly significant result when multiple independent trials are performed \citep{2010EPJC...70..525G}—but remains notable given the typical CHIME/FRB localization region, which has an effective radius of approximately 20 arcminutes \citep{Bhardwaj2021A,2023arXiv230705261L,2024ApJ...971L..51B}. As CHIME’s localization capabilities improve in the future with the availability of baseband localizations for the majority of CHIME bursts \citep{2023ApJ...950..134M} or through the use of CHIME/FRB outriggers \citep{2024AJ....168...87L}, we anticipate extending this analysis to larger distances. Finally, this distance limit is within the sensitivity range of current X-ray telescopes for detecting proposed FRB counterparts, such as SGR 1806–20-like magnetar giant flares \citep{2005Natur.434.1107P}, as discussed in \cite{2023arXiv230810930P}. Therefore, the selected galaxies would also be promising candidates for FRB multi-wavelength counterpart studies if an association were found. 

Further refining our selection, we imposed several criteria. Galaxies in the sample were required to have known SMs and SFRs, as listed in the HECATE catalog, and they must have non-zero visibility within their angular extent according to the CHIME/FRB Catalog-1 exposure map \citep{CHIMEFRBcatalog}. In this work, we define a galaxy's angular extent by assuming a circle centered at the optical centroid of the HECATE galaxies, with a radius equal to $2 \times$ the D25 isophotal radius along their major axis (hereafter D25). D25 is empirically found to be $\gtrsim 2 \times$ the effective (half-light) radius of disk galaxies \citep{1991ApJ...368...60P}. As a result, this angular extent is expected to capture over 99\% of the galaxy's stellar population. Note that this angular distance constraint, for example, would include FRB 20200120E source that is offset from M81 by 19.6' \citep{Kristen2022Nature} which is smaller than twice the M81's D25 (=21.6'). Also, if any of the Catalog-1 FRBs show a larger offset from the selected local volume galaxies, it would have been included in the FRB sample of \cite{2024ApJ...971L..51B} as the authors considered all FRBs in Catalog-1 with extragalactic dispersion measure (DM) $<$ 100 pc cm$^{-3}$ in their `low-DM' FRB sample. This is because the expected host and intergalactic medium contribution of the FRBs associated with the selected sample of galaxies would be insignificant at offsets $> 2 \times$ D25, as was seen in the case of FRB 20200120B where the estimated extragalactic DM is 40 pc cm$^{-3}$\citep{Bhardwaj2021A}. 

Finally, to mitigate observed selection bias against detecting FRBs in highly inclined galaxies, we adopted an additional constraint, selecting only galaxies with an inclination angle $i > 60^\circ$. Here, $i$ is defined such that $i = 0^\circ$ corresponds to a face-on galaxy and $i = 90^\circ$ to an edge-on system. Recent work by \citet{2024arXiv240801876B} demonstrates that FRBs are less likely to be detected in edge-on galaxies, likely due to propagation effects such as scattering along the disk plane. As this orientation-dependent bias can affect both detection efficiency and host association, we conservatively exclude low-inclination galaxies to avoid underestimating the expected number of detectable FRBs.
After applying all these criteria, our final sample consisted of 495 galaxies. The sky distribution of these selected galaxies is shown in Figure \ref{fig:frbfov}. Note that if we do not employ the inclination angle selection criterion, our local volume galaxy sample would consist of 814 galaxies\footnote{The local volume galaxy catalog of 814 galaxies, including physical parameters and CHIME/FRB exposure values, is available as a machine-readable table at \url{https://doi.org/10.5281/zenodo.15649649}.} which includes NGC 3252, the host galaxy of the second closest extragalactic FRB 20181030A \citep{Bhardwaj2021B}, an edge-on system. In \S\ref{subsec:gamma-r4}, we discuss the implications of excluding galaxies based on the inclination angle criterion.

\section{Analysis \& Results}
\label{sec:results}

Following the identification of the local volume galaxy sample in \S\ref{sec:LVsample}, we cross-matched these galaxies with Catalog-1 FRBs that have a signal-to-noise ratio (S/N) $>$ 12 as reported by \texttt{bonsai}, the real-time tree dedispersion algorithm used in the CHIME/FRB search pipeline \citep{2018ApJ...863...48C}. In this context, the S/N is defined as the maximum value of the dedispersed and boxcar-smoothed time series divided by the root-mean-square of the noise in the baseline. As noted by \cite{CHIMEFRBcatalog}, the reported fluence measurements in Catalog-1 are subject to uncertainty, we adopted the same framework as \cite{CHIMEFRBcatalog} and \cite{2023ApJ...944..105S}, using S/N as a proxy for fluence since it is strongly correlated \citep{2023AJ....165..152M}. We considered an FRB potentially associated with a galaxy if its 2$\sigma$ localization region overlapped with a circular region centered on the galaxy’s coordinates, with a radius of 2$\times$D25. 

First, independent of the S/N threshold, we examined the 62 bursts from 18 repeating FRB sources in Catalog-1. Each repeater had at least one burst meeting the S/N threshold. Using the published precise 2$\sigma$ localization regions of the 18 repeaters \citep{CHIMEFRB2019, 2020ApJ...891L...6F, 2020Natur.577..190M, Bhardwaj2021A, 2023ApJ...950..134M, 2024ApJ...971L..51B, 2024ApJ...961...99I}, we found no associations between these repeaters and the 495 selected local volume galaxies. Our analysis then focused on the remaining 474 non-repeating FRBs, of which only 317 have an S/N $> 12$. Of these 317 FRBs, baseband localization regions with sub-arcminute precision were available for 112 bursts \citep{2024ApJ...969..145C}. We did not find any associations between the 112 FRBs and the local volume galaxies. For the remaining 205 Catalog-1 bursts with S/N $>$ 12, we used the 95\% confidence header localization regions from Catalog-1. Again, no associations were identified.
Based on the injection-assisted fluence–S/N relation reported by \cite{CHIMEFRBcatalog}, we conclude that no Catalog-1 FRBs with fluence exceeding 5 Jy ms are likely to be associated with these galaxies. Our injection-determined understanding of sensitivity makes this conclusion highly plausible. In the future, we can further strengthen these results by using fluences derived from baseband data, which have considerably lower uncertainties than those based on Catalog-1 intensity measurements.

Given the lack of associations, we now constrain the \(\gamma\)-$E_{*}$ parameter space of the FRB energy function. Using the proportionality constants \( K_{0} \) estimated in \S\ref{sec:formalism} for different combinations of \(\gamma\) and $E_{*}$ (see Figure \ref{fig:Ko}), we calculate the expected number of FRBs (\( N_{\rm FRB} \)) from our sample of local volume galaxies during the duration of the Catalog-1 FRBs using Equation \ref{Eq1:gal_rate}. This calculation incorporates the SFR and SM values from the HECATE catalog alongside the mean exposure of the local volume galaxies in Catalog-1. Assuming that \( N_{\rm FRB} \) follows a Poisson distribution, we note that the total number of FRBs expected from 495 galaxies over the Catalog-1 observation period is a Poisson-distributed random variable. We then constrain the \(\gamma\)-$E_{*}$ parameter space using the fact that no FRBs were detected from the local volume galaxy sample, as shown in Figure \ref{fig:expected_frbs}. This allows us to estimate a 1$\sigma$ Poisson confidence upper limit on \(\gamma\) for a given $E_{\star}$, which is discussed in \S\ref{subsec:lower-limit}.

\subsection{Upper Limit Constraints on $\gamma$}
\label{subsec:lower-limit}

We estimate an upper limit on $\gamma$ by adopting a fixed value of $E_{*}$. Several studies have constrained $E_{*}$ using the full CHIME/FRB sample \citep{Hashimoto2022MNRAS,2023ApJ...944..105S}. In this work, we adopt $\log(E_{*}/\text{erg}) = 41.38$, as estimated by \cite{2023ApJ...944..105S}. This value is comparable to, though slightly lower than, those reported in other studies that analyzed FRBs detected at $\sim$1 GHz \citep{Luo2020MNRAS, 2022MNRAS.509.4775J}. 

\begin{figure}[htbp]
    \centering
    \includegraphics[width=0.99\textwidth]{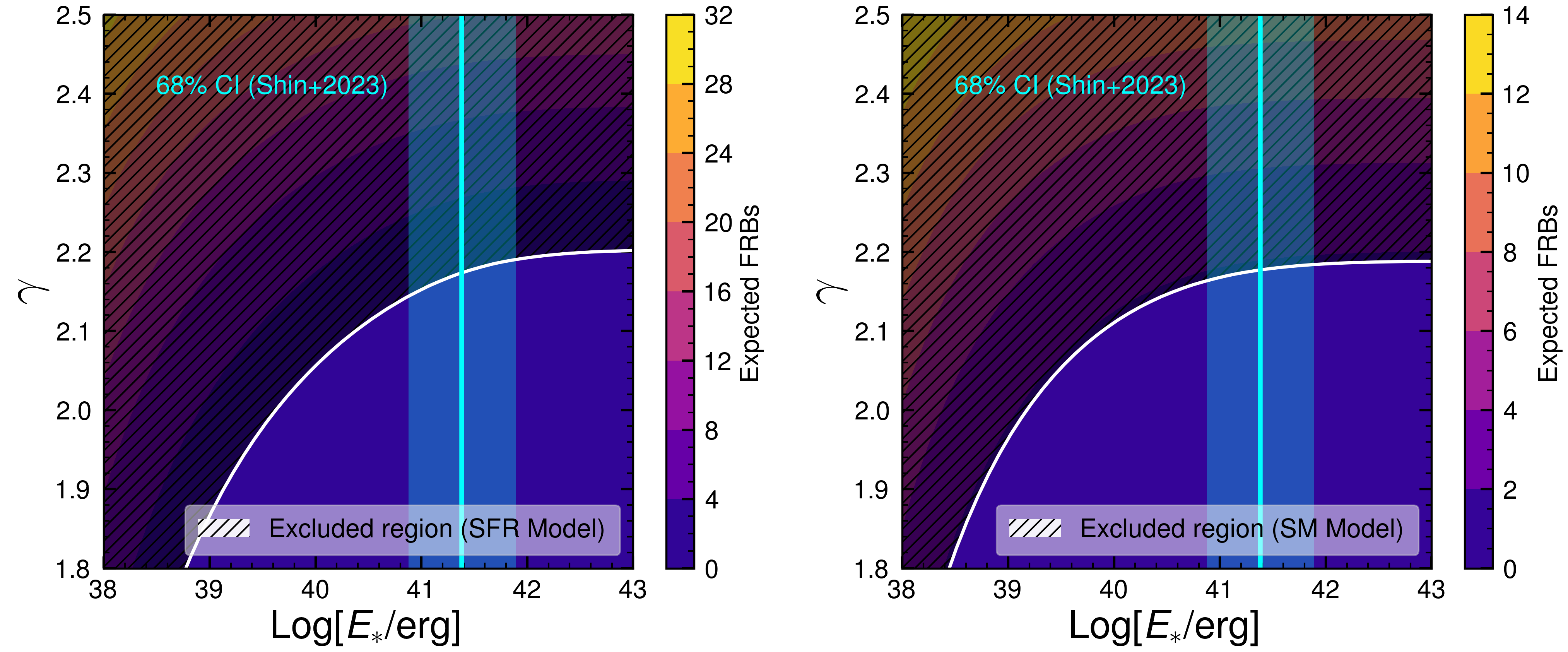}
    \caption{Predicted number of FRB detections as a function of $\gamma$ and $E_*$ for the SFR (left column) and SM (right column) models. The hatched region denotes the excluded $\gamma$-$E_*$ parameter space, constrained by the non-detection of FRBs from selected local volume galaxies in Catalog-1. The shaded vertical region represents the 1$\sigma$ credible interval for $E_*$ as derived by \cite{2023ApJ...944..105S}.}
    \label{fig:expected_frbs}
\end{figure}

Using the method described by \cite{1986ApJ...303..336G}, we compute the one-sided $1\sigma$ Poisson confidence upper limit on $\gamma$ at $\log(E_{*}/\text{erg}) = 41.38$ and find $\gamma < 2.18$ for both the SFR and SM models. The corresponding 95\% confidence upper limits are $\gamma < 2.3$ in both models. This constraint remains consistent across a range of $E_{*}$ values within their uncertainty limits (see Figure \ref{fig:expected_frbs}) and is robust against variations in $z_{\rm max}$ and $\R_{\text{all-sky}}$, as discussed in Appendix \ref{appendix:rate_zmax}. Thus, we conclude that if the FRB energy function described in \S\ref{sec:formalism} extends down to $E = 3 \times 10^{34}$ erg, then $\gamma$ should satisfy $\gamma < 2.3$. The threshold energy of $3 \times 10^{34}$ erg is comparable to the estimated isotropic-equivalent energy of the Galactic magnetar burst from SGR 1935+2154 \citep{CHIME2020Nature}, making it a physically motivated reference point for our analysis. However, we note that the constraint on $\gamma$ becomes significantly weaker for larger $E_{\rm th}$ values, as illustrated in Appendix~\ref{appendix:Eth}.

\subsection{Constraints on $\gamma$ using FRB 20181030A}
\label{subsec:gamma-r4}

As noted in \S\ref{sec:LVsample}, our local volume galaxy sample does not include the host galaxy of FRB 20181030A \citep{Bhardwaj2021B}, NGC 3252, due to the inclination angle cutoff criterion. If we do not consider this criterion in the sample selection, we obtain a sample of 814 galaxies within 21 Mpc, including NGC 3252. With this expanded sample, we followed the same procedure described above to identify potential associations with the bright Catalog-1 FRBs and found only FRB 20181030A. We can then constrain \(\gamma\) using the framework discussed in \S\ref{subsec:lower-limit}, based on the association of this single FRB with the expanded local volume galaxy sample. We derive constraints for both the SFR and SM models, resulting in \(\gamma = -2.0^{+0.2}_{-0.2}\) (1$\sigma$ credible region) using log($E_{*}$/erg) = 41.38. This constraint is consistent with the permitted region derived in \S\ref{subsec:lower-limit}
, albeit providing a more stringent lower limit on \(\gamma\). However, as the inclination angle bias makes our estimation more conservative, we primarily use the derived constraints on \(\gamma\) that include the inclination angle selection criterion for the discussion of this work's implications.

\section{Discussion}
\label{sec:discussion}

In \S\ref{sec:results}, we derived an upper limit on $\gamma$, the faint-end slope of the FRB energy function, using FRBs from the CHIME/FRB Catalog-1 and a well-defined sample of local volume galaxies. Our most conservative constraint, based on the SM model, yields $\gamma < 2.2$ at the $1\sigma$ confidence level, or $\gamma < 2.3$ at 95\% confidence upper limit, which serves as the basis for our subsequent discussion.

\begin{figure}[ht]
\begin{center}
\includegraphics[width=0.7\linewidth]{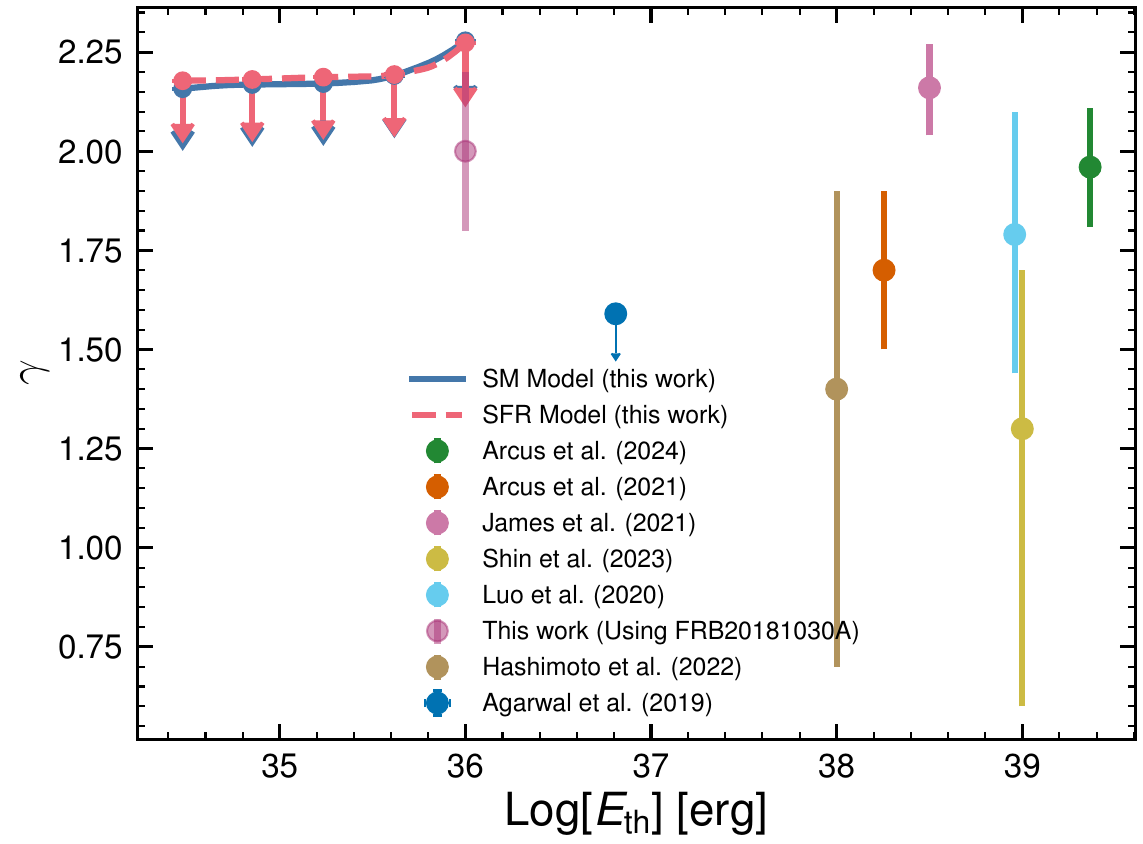}
\end{center}
\caption{Comparison of the constraints on the faint-end slope of the FRB energy function, $\gamma$, for various threshold energy ($E_{\rm th}$) values. In this work, we vary $E_{\rm th}$ between $10^{34}$ and $10^{36}$ ergs—corresponding to the energy of the burst detected by CHIME/FRB from SGR 1935+2154 and the isotropic energy of FRB 20181030A (given a fluence of 5 Jy ms), respectively (see \S\ref{sec:LVsample}). For limits corresponding to higher $E_{\rm th}$ values, see Appendix \ref{appendix:Eth}. Also shown are constraints reported in previous studies \citep{agrawal2019MNRAS, Luo2020MNRAS, Arcus2021MNRAS, Hashimoto2022MNRAS, 2020MNRAS.498.3927H, 2022MNRAS.509.4775J, 2023ApJ...944..105S}. Error bars represent the 1$\sigma$ error region.}
\label{fig:gamma_constraints}
\end{figure}

This analysis provides the most stringent constraint on \(\gamma\) at the lowest energy threshold to date, with \(E_{\rm th} = 3 \times 10^{34}\) erg. This threshold is an order of magnitude lower than those used in previous studies \citep{agrawal2019MNRAS,Luo2020MNRAS,Arcus2021MNRAS,Hashimoto2022MNRAS,2020MNRAS.498.3927H,2022MNRAS.509.4775J,2023ApJ...944..105S}. Despite the lower threshold, our constraints on \(\gamma\) are consistent with those derived at higher energy thresholds in earlier works, suggesting the possibility of a universal FRB energy function extending to these lower energies. Figure \ref{fig:gamma_constraints} compares our derived constraints on $\gamma$ with those from previous studies. 

Importantly, our constraint that $\gamma < 2.3$ suggests that the FRB population is dominated by intrinsically energetic bursts. A luminosity function with such a flat power-law index (i.e., $\gamma < 2.5$) implies that the increase in comoving volume with distance outweighs the decrease in observed flux density, leading to a detectable population primarily composed of distant, cosmological FRBs \citep{Macquart2018MNRAS}. This theoretical expectation directly leads to the observable fluence-extragalactic dispersion measure (F-DM) anticorrelation.

The F-DM anticorrelation is a key observational signature of a cosmological FRB population, where intrinsically bright bursts at greater distances (higher DMs) appear fainter (lower fluences). This relationship was first  observed at 1.4 GHz by \cite{Shannon2018Nature} (often referred to as the `dispersion-brightness relation'). Our finding of a flat luminosity function from local FRBs provides empirical support for these theoretical predictions, as a shallower energy function naturally results in a detectable population where the F-DM anticorrelation is pronounced. This is particularly relevant at higher redshifts where the intergalactic medium (IGM) contribution to FRB DMs dominates and the relative host galaxy DM contribution decreases \citep{2023MNRAS.518..539M}.

This study reinforces the role of DM as a useful distance proxy in studies of distant FRBs and strengthens the case for using them as cosmological probes \citep{Macquart2020Nature}. In the CHIME band (0.4--0.8 GHz), a mild (albeit low statistical significance; $p$-value = 0.02) F-DM anticorrelation has also been identified in a sample of 137 FRBs from the first CHIME/FRB baseband catalog \citep{2025ApJ...979..160S}. While a comprehensive re-analysis and plotting of the F-DM distribution for the full CHIME population is beyond the scope of this work, these published results further validate the expected relationship across different frequency bands.

Finally, as illustrated in Figure \ref{fig:expected_frbs}, the overlap in the constraints on $\gamma$ between the SFR and SM models precludes us from distinguishing between different FRB source formation scenarios based solely on comparisons with existing constraints on $\gamma$ at higher energies (see Figure \ref{fig:gamma_constraints}). Nonetheless, our results have important implications for the SFR model, particularly those involving young magnetars. 
As specifically illustrated in Figure 4 of \cite{Margalit2020}, a relatively flat power-law index for the differential luminosity function -- such as our constraint of $\gamma < 2.3$ (which corresponds to their integrated power-law index of $<1.3$) -- implies that the total observed FRB rates can be consistent with, and potentially dominated by, intrinsically less energetic sources. In this context, our derived $\gamma$ is consistent with models in which bursts from magnetars similar to SGR 1935+2154 could dominate the FRB population, provided their volumetric formation rate is sufficiently high and magnetar burst rates scale with age and magnetic field strength, as theorized in models used by \cite{Margalit2020}.

However, reconciling our inferred $\gamma < 2.3$ with the observed activity of hyperactive repeating FRBs—such as FRB~20121102A, FRB~20180916B, FRB~20201124A, FRB~20220912A, and FRB~20240114A \citep{2020Natur.582..351C,2021Natur.598..267L,2022Natur.606..873N,2022RAA....22l4002Z,2023arXiv230414671F,2023ApJ...955..142Z}—remains nontrivial. These sources exhibit burst rates and energetics that far exceed the behavior of typical Galactic magnetars like SGR~1935+2154 \citep{2023A&A...674A.223P}. As shown in Figure~5 of \citet{Margalit2020}, if such hyperactive FRBs originate from the same population as lower-activity sources, then a flat luminosity function with $\gamma < 2.3$ implies a volumetric formation rate of FRB-producing magnetars of $\lesssim 100$\,Gpc$^{-3}$\,yr$^{-1}$. This is in clear tension with the expected formation rate of SGR~1935+2154-like magnetars, which is $\gtrsim 10^4$\,Gpc$^{-3}$\,yr$^{-1}$ \citep[$\gtrsim 10\%$ of the core-collapse supernova rate;][]{2012ApJ...757...70D,2015SSRv..191..315M}. One possible resolution, as proposed by \cite{2025arXiv250609138B}, is that FRBs arise from a single population with a broad power-law distribution in intrinsic activity rates, such that hyperactive repeaters represent the rare, high-activity tail. This framework avoids the need to invoke multiple progenitor classes and remains consistent with our $\gamma$ constraint. Nonetheless, the marked disparity in observed repetition rates still leaves open the possibility that a subset of FRBs—particularly the most active ones—may originate from distinct formation channels. Thus, while a universal energy function for all FRBs remains plausible, it is not yet conclusively established, and future studies must account for both the statistical distribution of burst activity and the physical diversity of progenitor environments \citep{2023PASA...40...57J}.

 \section{Conclusion}
\label{sec:conclude}

We conducted a systematic search for associations between CHIME/FRB Catalog-1 bursts and local volume galaxies within a luminosity distance of 21 Mpc to constrain the faint-end slope, $\gamma$, of the FRB energy function, assuming a Schechter-like distribution. Based on the absence of bursts from 495 local volume galaxies, we estimate the 1$\sigma$ single-sided confidence upper limit on $\gamma$ to be $<$ 2.2, or $\gamma < 2.3$ at 95\% confidence upper limit. Our findings suggest that the FRB population is dominated by bright, cosmological bursts, with the energy function extending into lower energy regimes than previously constrained. Moreover, if we ignore the effect of an host inclination-related selection bias, we derived constraints for both the SFR and SM models, resulting in \(\gamma = -2.0^{+0.2}_{-0.2}\) (1$\sigma$ credible region). However, if the FRB threshold energy exceeds $10^{36}$ ergs \citep[for example, see][]{2025PASA...42...17H}, our upper limit becomes less constraining (see Appendix \ref{appendix:Eth}).

Our constraint on $\gamma$ reinforces that the observed FRB population is dominated by intrinsically energetic bursts. However, theoretical interpretations of this result diverge. In particular, \citet{Margalit2020} argue that reproducing the observed repetition rates of hyperactive sources under such a flat energy distribution requires a separate, rare population of more extreme magnetars formed through non-standard channels. By contrast, \citet{2025arXiv250609138B} demonstrate that a single population of magnetars with a broad power-law distribution in intrinsic activity rates can reproduce both one-off and repeating FRBs within a unified framework. Our results are therefore consistent with both scenarios but highlight a tension between these competing interpretations. Future observations that better characterize repetition rates and host environments across the FRB population will be essential to distinguishing between these possibilities.

Our study underscores the effectiveness of CHIME/FRB’s wide field of view and high sensitivity in constraining the faint-end of the FRB luminosity function through observations of local volume galaxies. These galaxies offer critical opportunities for multi-wavelength follow-up studies, which are instrumental in probing the nature of FRB progenitors \citep{Bhardwaj2021B,2023arXiv230810930P,2024ApJ...971L..51B}.

Looking ahead, the forthcoming CHIME/FRB Catalog-2, which extends the observation period significantly, is poised to provide more stringent constraints on $\gamma$ and further test the universality of the FRB energy function. The expanded dataset is expected to enhance the detection of fainter FRBs and improve our understanding of their energy distribution. Continued observations, in conjunction with deeper and more sensitive surveys, will be essential for refining these constraints and advancing our comprehension of FRB origins and their broader astrophysical implications.

\begin{acknowledgments}
We thank the anonymous reviewer for their careful reading of our manuscript and their insightful comments and suggestions.
We acknowledge that CHIME is located on the traditional, ancestral, and unceded territory of the Syilx/Okanagan people. We are grateful to the staff of the Dominion Radio Astrophysical Observatory, which is operated by the National Research Council of Canada.  CHIME is funded by a grant from the Canada Foundation for Innovation (CFI) 2012 Leading Edge Fund (Project 31170) and by contributions from the provinces of British Columbia, Qu\'{e}bec and Ontario. The CHIME/FRB Project is funded by a grant from the CFI 2015 Innovation Fund (Project 33213) and by contributions from the provinces of British Columbia and Qu\'{e}bec, and by the Dunlap Institute for Astronomy and Astrophysics at the University of Toronto. Additional support was provided by the Canadian Institute for Advanced Research (CIFAR), the Trottier Space Institute at McGill University, and the University of British Columbia.

M.B is a McWilliams fellow and an International Astronomical Union Gruber fellow. M.B. also receives support from the McWilliams seed grant. V.M.K. holds the Lorne Trottier Chair in Astrophysics \& Cosmology, a Distinguished James McGill Professorship, and receives support from an NSERC Discovery grant (RGPIN 228738-13). K.W.M. holds the Adam J. Burgasser Chair in Astrophysics and is supported by an NSF grant (2018490). M.W.S. acknowledges support from the Trottier Space Institute Fellowship program. 

\end{acknowledgments}

\bibliographystyle{aasjournal}
\bibliography{sample631.bib}
\appendix

\section{Stellar Mass Density Derivation}
\label{appendix:sm}
The stellar mass density, $\rho_*(z)$, represents the total mass density of long-lived stars and stellar remnants that has accumulated from earlier episodes of star formation. For a given redshift, $z$, $\rho_*(z)$ can be derived from the cosmic star formation rate density, $\psi(z)$, by integrating over all prior epochs:
\begin{equation}
\rho_*(z) = (1 - R) \int_{z}^{\infty} \frac{\psi(z')}{(1 + z') H(z')} \, dz',
\end{equation}
where $H(z)$ is the Hubble parameter at redshift $z$, and $R$ is the ``return fraction," defined as the mass fraction of each generation of stars that is returned to the interstellar medium (ISM) and IGM through processes such as supernovae and stellar winds. The remaining fraction, $(1 - R)$, represents the mass that remains locked in long-lived stars and stellar remnants. We assume $R = 0.27$ as prescribed by \cite{2014ARA&A..52..415M}.

For $\psi(z)$, we adopt the best-fitted model from \cite{2014ARA&A..52..415M}, given by:
\begin{equation}
\label{appendix:EqA2_SFRD}
\psi(z) = 0.015 \, \frac{(1 + z)^{2.7}}{1 + \left(\frac{1 + z}{2.9}\right)^{5.6}} \, M_\odot \, \mathrm{yr}^{-1} \, \mathrm{Mpc}^{-3},
\end{equation}
By integrating $\psi(z)$ from redshift $z$ to infinity, we obtain the total stellar mass density $\rho_*(z)$ at the redshift of interest.

\section{Impact of Redshift Threshold and All-Sky Rate on $\gamma$}
\label{appendix:rate_zmax}

In this section, we assess the sensitivity of our results to variations in the maximum redshift threshold, $z_{\text{max}}$, and the CHIME/FRB all-sky rate, $\R_{\text{all-sky}}$. Our primary constraint on the faint-end power-law index of the energy function, $\gamma$, are derived using $z_{\text{max}} = 2$ and $\R_{\text{all-sky}} = 525$ day$^{-1}$, as noted in \S\ref{sec:formalism}. However, the reported CHIME/FRB rate, including statistical and systematic uncertainties, is $[525 \pm 30\, (\text{stat.})^{+142}_{-131}\, (\text{sys.})]$ bursts per year.

\begin{figure}[ht]%
    \centering
    \includegraphics[origin=top, width=.95\linewidth]{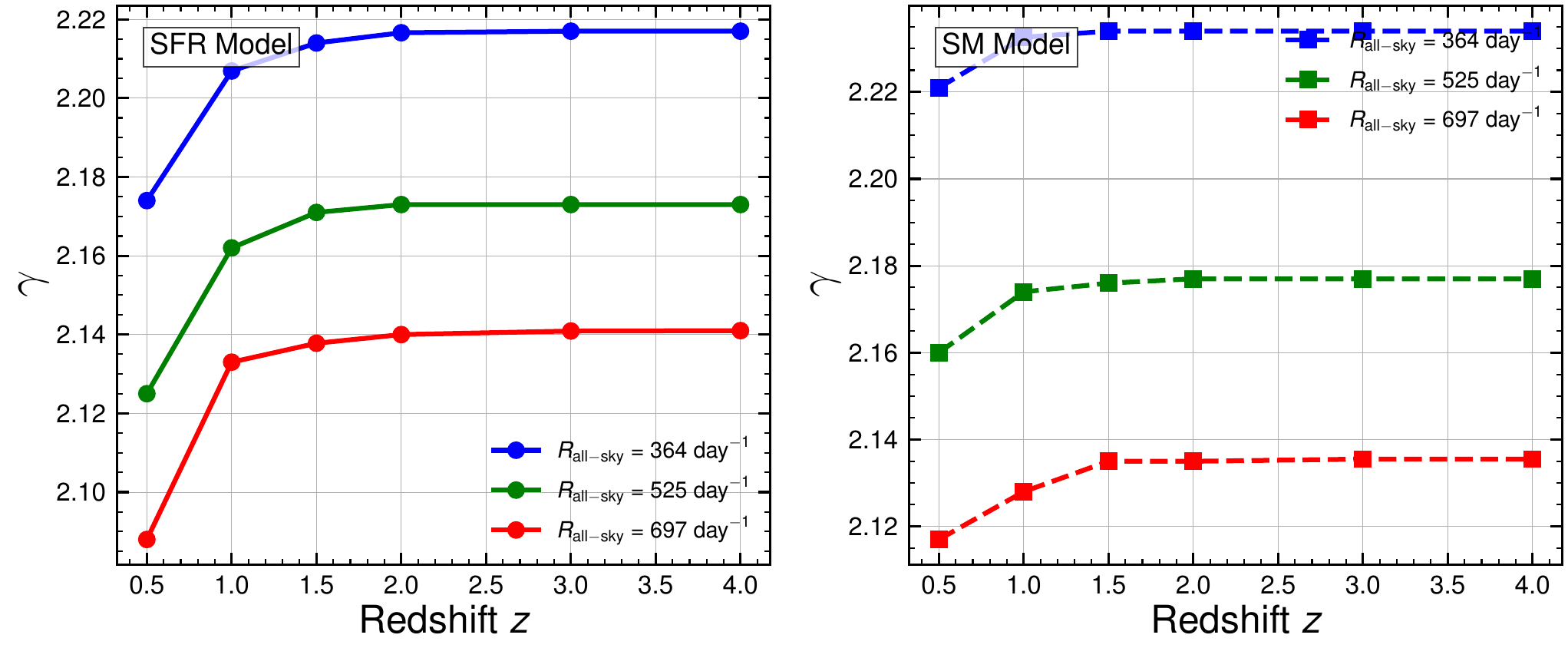}
    \caption{Derived 1$\sigma$ confidence upper limit on $\gamma$, for both the SFR and SM models, as a function of z$_{\text{max}}$ for different CHIME/FRB all-sky rate estimates ($\R_{\text{all-sky}}$).}
    \label{fig:three_images}
\end{figure}

To evaluate the impact of $\R_{\text{all-sky}}$ on our results, we adopt alternative values of $364$ day$^{-1}$ and $697$ day$^{-1}$, corresponding to the $1\sigma$ lower and upper limits after incorporating both systematic and statistical uncertainties. The resulting constraints on $\gamma$ exhibit minimal change ($|\Delta\gamma| \lesssim 0.08$) for both the SFR and SM models at a fixed $z_{\text{max}}$. This is illustrated in Figure \ref{fig:three_images}, where we plot the $1\sigma$ confidence upper limit on $\gamma$ assuming $\log(E_*/\text{erg}) = 41.38$.

Similarly, varying $z_{\text{max}}$ from 0.5 to 4.0 or beyond results in a negligible shift in $\gamma$ ($|\Delta\gamma| \lesssim 0.04$). This behavior is expected, as the upper limit on $\gamma$ is primarily constrained by the detection of low-energy FRBs, which remains largely unaffected by changes in $z_{\text{max}}$. The detectability of such bursts is predominantly determined by telescope sensitivity rather than the adopted redshift threshold (see Equation \ref{Eq:Elim}).

Thus, we conclude that our constraint on $\gamma$ is robust, and variations in either $z_{\text{max}}$ or $\R_{\text{all-sky}}$ within the explored ranges do not impact the overall conclusions of this study.

\section{Dependence of $\gamma$ on the Spectral Index $\alpha$}
\label{appendix:alpha}

In this section, we assess the sensitivity of our constraint on the faint-end slope $\gamma$ to the assumed spectral index $\alpha$ used when converting observed fluence to intrinsic energy (see Equation~\ref{Eq:Elim}). We vary $\alpha$ from 0 to $-2$, a range that encompasses both flat and steep spectra consistent with FRB observations. For each value of $\alpha$, we recompute the energy threshold $E_{\rm lim}$ and derive the corresponding $1\sigma$ upper limit on $\gamma$, holding all other model parameters fixed to their fiducial values described in Section~\ref{sec:formalism}.

\begin{figure}[ht]
    \centering
    \includegraphics[width=0.85\textwidth]{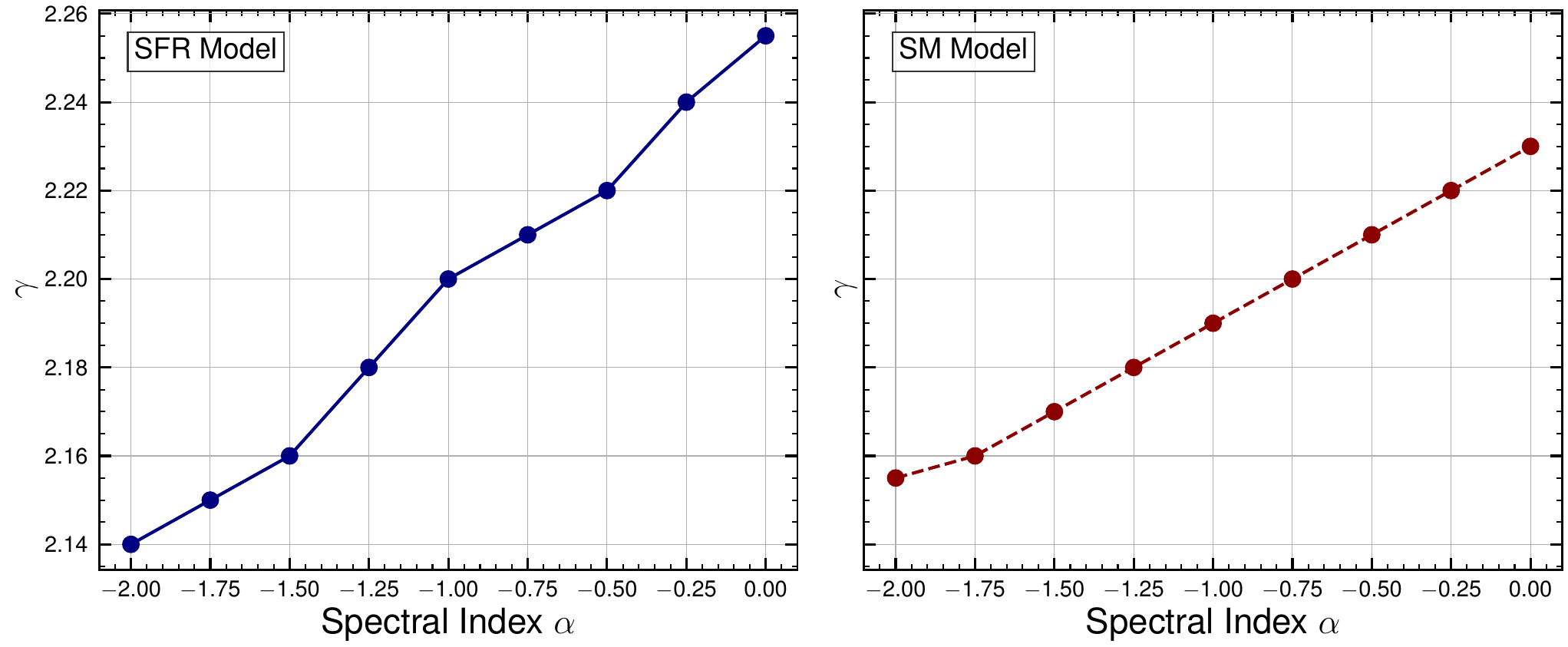}
    \caption{Derived $1\sigma$ confidence upper limits on the faint-end slope $\gamma$ as a function of the spectral index $\alpha$. Results are shown separately for the SFR model (solid line; left) and the SM model (dashed line; right).}
    \label{fig:alpha_comparison}
\end{figure}

Figure~\ref{fig:alpha_comparison} shows the resulting upper limits on $\gamma$ for both the SFR and SM models. The constraints change only weakly with $\alpha$, varying by less than 0.1 over the entire range. This behavior is expected, as the detectability of low-energy bursts—key to constraining the faint-end of the energy function—is dominated by low-redshift galaxies. In this regime, the redshift dependence in the fluence-to-energy conversion becomes insignificant, and hence variations in $\alpha$ do not significantly affect the inferred energy threshold.

\section{Dependence of $\gamma$ on $E_{\rm th}$}
\label{appendix:Eth}

In the absence of robust constraints on the minimum intrinsic energy of FRBs, we adopt $E_{\rm th} = 3 \times 10^{34} \, \mathrm{erg}$. This threshold corresponds to the isotropic energy of the bright CHIME/FRB burst from SGR~1935+2154 \citep{2020CHIMEFRB} and of the bursts detected by CHIME/FRB from the nearby repeating FRB~20200120E \citep{Bhardwaj2021A}. Raising $E_{\rm th}$ to $3 \times 10^{35} \, \mathrm{erg}$ accommodates the lowest-energy bursts observed from FRB~20181030A and FRB~20180916B, two local universe repeaters localized to spiral galaxies at distances of $\sim 21 \, \mathrm{Mpc}$ and $149 \, \mathrm{Mpc}$, respectively \citep{Bhardwaj2021B,2020Natur.577..190M}. Moreover, for the median distance of 4.2 Mpc galaxies in our sample, the isotropic energy for a putative burst with fluence of 5 Jy ms and bandwidth of 0.4 GHz corresponds to $E_{\rm th}$ = $4 \times 10^{34}$ erg, which is close to the aforementioned fiducial E$_{\rm th}$ value.

\begin{figure}[ht]%
    \centering
    \includegraphics[origin=top, width=.95\linewidth]{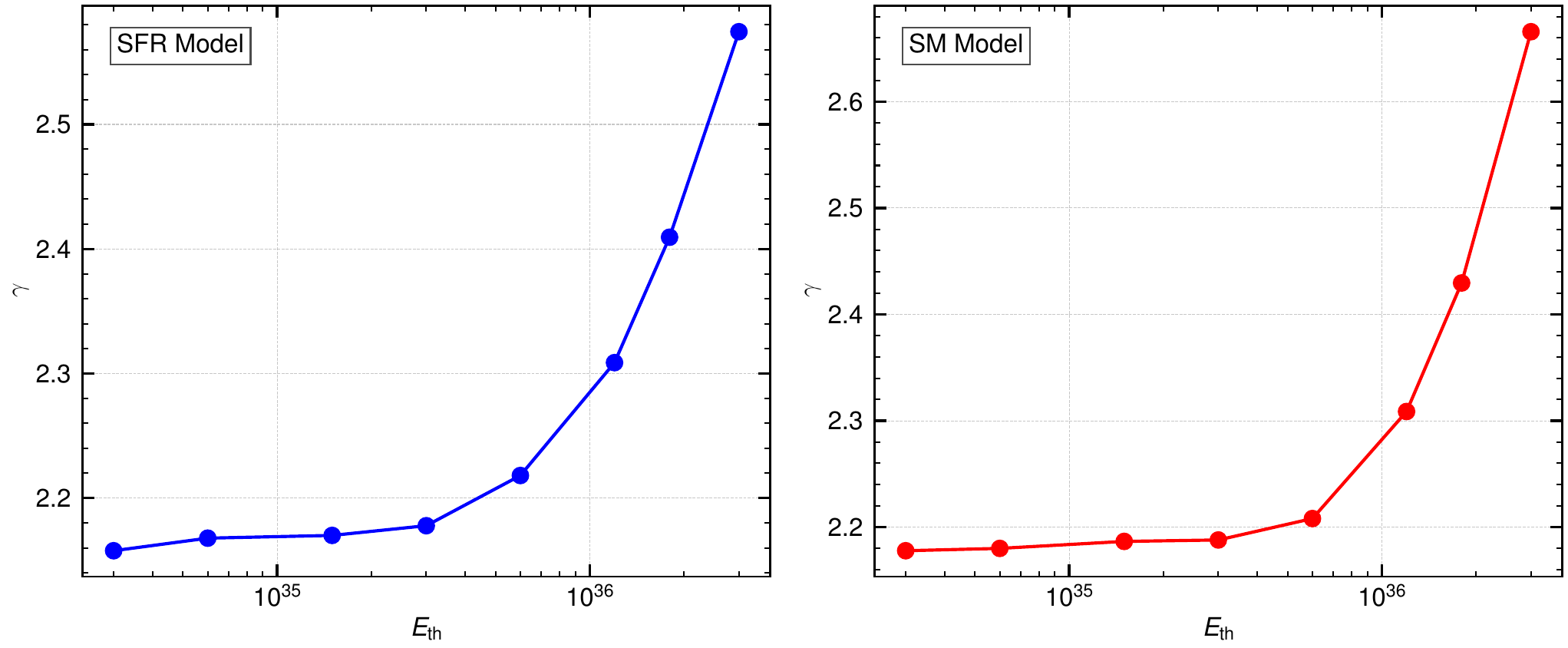}
    \caption{Derived 1$\sigma$ confidence upper limit on $\gamma$ for both the SFR and SM models across different threshold energy values (E$_{\rm th}$).}
    \label{fig:Eth_comparison}%
\end{figure}

Figure \ref{fig:Eth_comparison} illustrates the effect of varying the threshold energy ($E_{\rm th}$) on the upper limit of $\gamma$, while maintaining other parameters to their default values, as described in Section \ref{sec:formalism}. The constraints on $\gamma$ remain largely insensitive to plausible $E_{\rm th}$ values when $E_{\rm th}$ $< 10^{36}$ erg. However, increasing $E_{\rm th}$ beyond $3 \times 10^{36} \, \mathrm{erg}$ significantly weakens the constraints, limiting the utility of local volume galaxies in constraining the FRB energy function. 

\section{Impact of the 21 Mpc Distance Cutoff on the derived Rate Constraints}
\label{appendix:sim}
In this work, we constrain the FRB volumetric rate using the comoving volume defined by the closest repeating CHIME FRB in Catalog-1 (21 Mpc). To evaluate whether this a posteriori selection introduces bias, we performed a simulation assuming a true FRB rate of one event per unit time per unit volume. In one million simulated surveys—each containing 525 FRBs, consistent with Catalog-1—we generated event distances and identified the closest detection. The comoving volume enclosed within this distance, denoted as \(V_1\), defines the effective survey volume.

Using the 1\(\sigma\) lower and upper limit factors from \cite{1986ApJ...303..336G} for a single detection (as applied in Section~\ref{subsec:gamma-r4}), we constructed one-sigma confidence intervals for the volumetric rate, with limits of \(0.173/V_1\) and \(3.300/V_1\), respectively. Our simulation shows that the true rate is contained within these limits in approximately 80.5\% of cases, confirming that our method is conservative. Thus, the 21 Mpc cutoff naturally reflects the statistical variability inherent in a Poisson process and does not bias our volumetric rate constraint.
\end{document}